# Physical Characterisation of low delta-V asteroid (175706) 1996 FG3

Stephen D. Wolters[a,1], Ben Rozitis[a], Samuel R. Duddy[b], Stephen C. Lowry[b], Simon F. Green[a], Colin Snodgrass[c], Olivier R. Hainaut[d], and Paul Weissman[e]

[a] *Planetary and Space Sciences Research Institute, The Open University, Walton Hall, Milton Keynes, MK7 6AA, UK*
[b] *Centre for Astrophysics and Planetary Sciences, The University of Kent, Canterbury, Kent, CT2 7NZ, UK*
[c] *Max-Planck-Institut für Sonnensystemforschung, 37191 Katlenburg-Lindau, Germany*
[d] *European Southern Observatory, Karl-Schwarzschild-Str. 2, D-85748 Garching bei München, Germany.*
[e] *Planetary Ices Section, Jet Propulsion Laboratory, Pasadena, California, USA*

No. of manuscript pages: 24
No. of Figures: 7
No. of Tables: 8

e-mail: stephen.wolters@jpl.nasa.gov

e-mail addresses of co-authors: b.rozitis@open.ac.uk, s.duddy@kent.ac.uk, s.c.lowry@kent.ac.uk, s.f.green@open.ac.uk, snodgrass@mps.mpg.de, ohainaut@eso.org, paul.r.weissman@jpl.nasa.gov

---

[1] Current affiliation: Planetary Ices Section, Jet Propulsion Laboratory, 4800 Oak Grove Drive, Pasadena CA 91109, USA, stephen.wolters@jpl.nasa.gov, +1 (818) 454-0574

**Abstract**

Asteroid (175706) 1996 FG3 is a binary asteroid and the baseline target for the proposed MarcoPolo-R sample-return mission. We present thermal IR photometry obtained with the ESO VLT+VISIR together with optical photometry obtained with the ESO NTT+EFOSC2[2]. An absolute visual magnitude $H_V$ = 17.833 ± 0.024 and phase parameter $G$ = -0.041 ± 0.005 is derived. The Near-Earth Asteroid Thermal Model (NEATM) has been fitted to the measured fluxes to derive a geometric visual albedo $p_v$ = 0.046 ± 0.014, effective diameter at the observed aspect $D_{\text{eff}}$ = 1.68 ± 0.25 km, and beaming parameter $\eta$ = 1.15 for phase angle $\alpha$ = 11.7°. The Advanced Thermophysical Model (ATPM) has been fitted to the measured fluxes to derive a more accurate effective diameter $D_{\text{eff}}$ = 1.71 ± 0.07 km and albedo $p_v$ = 0.044 ± 0.004. Based on the ATPM results, assuming the same albedo for primary and secondary, we derive a primary mean spherical diameter $D_p$ = (1.69 $^{+0.18}/_{-0.12}$) km, secondary diameter $D_s$ = 0.51 ± 0.03 km, and a secondary orbital semi-major axis $a$ = (2.8 $^{+1.7}/_{-0.7}$) km. A low surface thermal inertia $\Gamma$ = 120 ± 50 J m$^{-2}$ s$^{-1/2}$ K$^{-1}$ was also derived, suggesting a dusty surface and raising questions as to the binary formation mechanism of this asteroid. These physical properties are used to predict a Yarkovsky drift in semi-major axis of (-60 $^{+31}/_{-45}$) m yr$^{-1}$.

**Keywords:**
Minor planets, asteroids; surveys; Infrared: Solar system.

# 1 Introduction

Near-Earth Asteroid (NEA) (175706) 1996 FG3, hereafter referred to as 1996 FG3, is the baseline target for proposed sample return mission MarcoPolo-R (Barucci et al., submitted to Experimental Astronomy). A Hohmann-like transfer orbit from the Earth only requires $\Delta v$ = 5.16 km s$^{-1}$ (Perozzi, Rossi & Valsecchi 2001; Christou 2003; Binzel et al. 2004). 1996 FG3 was discovered on 24 March 1996 by R. H. McNaught from the Siding Spring Observatory, New South Wales, Australia.

## 1.1 Previous Observations

1996 FG3 was found to be a binary asteroid from lightcurve observations (Mottola & Lahulla, 1998; Pravec et al. 1998; Pravec et al. 2000), and subsequently modelled in detail (Mottola & Lahulla 2000; Pravec et al. 2006; Scheirich & Pravec, 2009). Two plausible formation mechanisms for binary NEAs are rotational fission caused by the spin up of the NEA (Walsh, Richardson & Michel 2008) via the YORP (Yarkovsky O'Keefe Radzievskii Paddack) effect (Rubincam 2000) or tidal disruption (e.g. Walsh & Richardson 2008, and references therein). Whiteley (2001), Binzel et al. (2001), Somers et al. (2010) and de León et al. (2011) classified 1996 FG3 as a C-type. As noted in Morbidelli et al (2006), among all known C-type binaries, 1996 FG3 encounters the Earth with the lowest relative velocity (10.7 km s$^{-1}$). Because tidal disruption is more likely at low relative velocity, this property makes 1996 FG3 one of the best candidates for having been formed in recent times as a result of this mechanism.

Mottola & Lahulla (2000) observed mutual eclipse events in the system and found that the model which best fits their observations is a satellite with orbital semi-major axis $a$ = (1.7 ± 0.3) $D_p$, where $D_p$ is the diameter of the primary, and an orbital eccentricity $e$ = 0.05 ± 0.05. Their lightcurve yields a primary that is almost spherical, with normalized dimension axes of $A_p$ = 1.05 ± 0.02, $B_p$ = 0.95 ± 0.02 and $C_p$ = 0.70 ± 0.10 and an orbital period of the secondary $P_{\text{orb}}^{\text{sid}}$ = 16.135 ± 0.005 h. They determined an average bulk density of 1.4 ± 0.3 g cm$^{-3}$ for the system.

Pravec et al. (2000) found $H_V$ = 17.76 ± 0.03, $G_R$ = -0.07 ± 0.02, $V - R$ = 0.380±0.003, rotational period of the primary $P_p$ = 3.5942 ± 0.0002 h, and the same sidereal orbital period of the secondary as Mottola & Lahulla. Pravec at al. (2006) found that while the secondary's rotation period could not be established from the available data, an assumption of its rotation being synchronous with the orbital motion appeared plausible and a search in vicinity of $P_{\text{orb}}^{\text{sid}}$ provided an estimate of $P_s$ =16.15 ± 0.01 h. The data suggested that the secondary is moderately elongated with a lightcurve amplitude (corrected to exclude contributions of light from the primary) of ≈ 0.4 mag. with its long axis approximately aligned with the line connecting the centres of the bodies, as indicated by the secondary rotation component's minima occurring around the times of the mutual events.

Sheirich & Pravec (2009) modelled the binary system with a numerical model that inverts the long-period components of observed lightcurves. In contrast to Motolla & Lahulla, a rotationally symmetric shape is used for the primary by assuming it to be an oblate spheroid, with a spin axis normal to the orbital plane of the secondary, while the short-period component is subtracted from the lightcurve. The shape of the secondary is modelled as a prolate spheroid synchronously rotating so that its long axis is aligned with the line connecting the centres of the two bodies when at the pericentre. They found (3σ uncertainties) $D_s/D_p$ = 0.28 ($^{+0.01}/_{-0.02}$), ratio of semi-major axis of mutual orbit with equatorial semi-axis of the primary $a/A_p$ = 3.1 ($^{+0.9}/_{-0.5}$), pole ecliptic longitude $\lambda_p$ = 242° ± 96°, pole ecliptic latitude $\beta_p$ = -84° ($^{+14°}/_{-15°}$), $P_{orb}^{sid}$ = 16.14 ± 0.01 h, $e$ = 0.10 ($^{+0.12}/_{-0.10}$), ratio of the equatorial semi-axis of the primary with the secondary $A_p/A_s$ = 0.33 ($^{+0.07}/_{-0.08}$), ratio of the equatorial and polar semi-axis of the primary $A_p/C_p$ = 1.2 ($^{+0.5}/_{-0.2}$), and ratio of the equatorial and polar semi-axis of the secondary $A_s/C_s$ = 1.4 ($^{+0.3}/_{-0.2}$). They also determined an average bulk density of (1.4 $^{+1.5}/_{-0.6}$) g cm$^{-3}$ for the system.

P. Pravec and collaborators are in the process of deriving an updated orbital model incorporating new photometric observations, including some taken in the weeks surrounding ours (Scheirich et al., 2011). Pravec (personal communication) has been able to determine that the January 2011 VLT and NTT observations presented in this paper were taken outside mutual events, while the March 2011 NTT observations were taken almost completely in an eclipse between the components of the system.

As stated above, both Mottola & Lahulla (2000) and Sheirich & Pravec (2009) determined a system average bulk density of ~1.4 g cm$^{-3}$. This bulk density is consistent with the average bulk density associated with C-type asteroids (Britt et al. 2002). Its low value is also highly suggestive of a rubble pile structure, making both YORP-induced rotational fission and tidal disruption viable formation mechanisms for the binary.

**1.2 Thermal modelling**

An asteroid's effective diameter [the equivalent diameter of a sphere with the same projected area as the (generally) irregularly shaped asteroid] can be related to its absolute visual magnitude $H_V$ and geometric albedo $p_v$ by (e.g. Fowler & Chillemi 1992):

$$D_{eff}(km) = \frac{10^{-H_V/5} \, 1329}{\sqrt{p_v}} \quad (1)$$

If we can measure both the scattered sunlight at an asteroid's surface and the absorbed and re-emitted thermal IR flux, a unique albedo and diameter can be derived. However we cannot directly measure the total radiation emitted in all directions. Instead a thermal model is used to simulate a surface temperature distribution. We used three simple thermal models: the Near-Earth Asteroid Thermal Model (NEATM, Harris 1998), the Standard Thermal Model (STM, Lebofsky et al. 1986) and the Fast Rotating Model (FRM, Lebofsky & Spencer 1989), as well as the more sophisticated Advanced Thermophyiscal Model (ATPM, Rozitis & Green 2011). The Planck function is numerically integrated over the visible hemisphere to provide a model IR flux $F_{MOD}(\lambda_n)$ that can be best-fit to the observed fluxes $F_{OBS}(\lambda_n)$.

*Simple Thermal Models*

In the NEATM, the maximum temperature ($T_{max}$) is given by:

$$T_{\max} = \left[\frac{(1-A_B)S}{\eta\varepsilon\sigma}\right]^{\frac{1}{4}} \qquad (2)$$

where $A$ is the bolometric Bond albedo, $S$ is the incident solar flux, $\eta$ is the so-called "beaming parameter", $\varepsilon$ is the thermal IR emissivity (0.9 is assumed) and $\sigma$ is the Stefan-Boltzmann constant. NEATM models the asteroid as a sphere and calculates the temperature on the surface assuming Lambertian emission on the day side and zero emission on the night side. The NEATM allows $\eta$ to be varied until $F_{\text{MOD}}(\lambda_n)$ gives a best fit to the observed thermal IR spectrum $F_{\text{OBS}}(\lambda_n)$, effectively forcing the model temperature distribution to show a colour temperature consistent with the apparent colour temperature implied by the data.

The STM describes an end case where the asteroid surface is in instantaneous equilibrium with the solar radiation, such as would be expected for a slow rotator with low surface thermal inertia, while the FRM assumes an iso-latitudinal temperature distribution suitable for a fast rotating high thermal inertia surface. Fits to these simple thermal models can be used to delineate a range of plausible surfaces.

*Advanced Thermophysical Model*

Thermophysical models combine detailed shape models with sophisticated thermal physics in order to determine accurate surface and sub-surface temperature distributions, and thermal emission fluxes. They attempt to account for all properties and processes that affect the observed asteroid thermal emission. In particular, they take into account shadowing and self-irradiation of non-convex bodies, heat conduction caused by non-zero surface thermal inertia, and thermal infrared beaming that is induced by unresolved surface roughness. Thermophysical models are used to gain further insights on physical properties and increased accuracy over simple thermal models and are also used when simple thermal models are inadequate in accurately reproducing certain observations. Furthermore, the asteroid Yarkovsky orbital drift and YORP rotational acceleration (Bottke et al. 2006, and references therein), which are caused by the net force and torque resulting from the asymmetric reflection and thermal re-radiation of sunlight from an asteroid's surface, can be predicted by thermophysical models. The ATPM used here is the first thermophysical model that includes thermal infrared beaming due to surface roughness when calculating Yarkovsky forces and YORP torques. This has been shown to increase the Yarkovsky orbital drift by up to 100% and dampen the YORP rotational acceleration by up to 50% (Rozitis & Green 2010, submitted to MNRAS).

## 2. Observations and Data Reduction

1996 FG3 was observed at the European Southern Observatory Very Large Telescope Unit 3 "Melipal" (ESO VLT UT3) at Paranal, Chile on 19.2 Jan 2011 UTC using the VISIR mid-IR instrument (Lagage et al. 2004) in imaging mode and at the ESO New Technology Telescope at La Silla on 29.3 Jan 2011 and 6.1 Mar 2011 UTC using the EFSOC2 instrument. The geometric circumstances are given in Table 1.

**Table 1: Observation Geometry**

| Telescope/ Instrument | Date of observation (UTC) | Heliocentric distance r (AU) | Geocentric distance Δ (AU) | Solar phase angle α (°) |
|---|---|---|---|---|
| VLT/VISIR | 2011-01-19 03:32-06:13 | 1.377 | 0.4047 | 11.7 |
| NTT/EFOSC2 | 2011-01-29 05:31-06:26 | 1.396 | 0.4114 | 1.4 |
| NTT/EFOSC2 | 2011-03-06 00:13-03:40 | 1.423 | 0.592 | 34.3 |

**2.1 Optical photometry**

The observations and data analysis of each observing run were conducted in a similar manner. The January NTT observations were timed such that the rotational phase covered by the optical observations would match the rotational phase covered by the thermal observations taken ~ 10 days previously. The March observations were conducted to extend the phase angle coverage and to contribute to a new determination of the $H,G$ parameters. The EFOSC-2 instrument at the NTT was used in imaging mode using the R-filter (Bessel). The 2048×2048 pixel CCD specification is given at: http://www.eso.org/sci/facilities/lasilla/instruments/efosc/inst/Ccd40.html.
Observations were conducted using exposure times of 60s, employing 2 × 2 pixel binning, providing a pixel scale of 0.24 arcsec per pixel and a field of view of approximately 4 × 4 arcmin. This provided an adequate number of comparison stars for differential photometry. To increase the S/N of the NEA signal, differential tracking at the rate of the NEA was used. However, the exposure time of each observation was short enough to keep stellar trailing within the seeing disc.

The data were first bias-corrected using a combination of bias frames taken throughout the night, and an overscan region in each frame. Flat fielding was performed using the median of a series of twilight sky flats, and finally a fringe map correction was applied. The images were then registered and shifted producing a set of images in which the stars were aligned in the field of view. The resulting images were then shifted again according to the rate of motion of 1996 FG3, producing a set of images in which the asteroid appeared stationary in the field of view. Images in each image set were then co-added in groups to increase the S/N of the measured NEA brightness and thus the fidelity of the extracted lightcurves.

Aperture photometry was performed on both coadded image sets using a range of aperture radii 1-5 FWHM (1.5-7.5 arcsec). Relative photometry was then performed using a composite of several comparison stars, which were checked for variability. A plot of brightness versus aperture radius showed that the background sky was not adversely influencing the photometry of the NEA and thus aperture corrections were not employed in determining its apparent magnitude. The approximate mean magnitude of the January lightcurve, estimated from a comparison of the observed partial lightcurve with the published lightcurves referenced in Section 1.1, was used to determine the apparent magnitude of the NEA at the time of observation. Pravec (personal communication) has independently derived an offset for our January observations and found that they agree with our estimates within 0.001 mag. The extinction coefficient was measured directly using the comparison stars in the field of view while standard star fields, chosen from Landolt (1992), were observed to derive the instrumental zero point. Calibration frames were obtained using the 0.6m telescope at the Table Mountain Observatory (Wrightwood, CA) to ensure the reliability of these values. The relative lightcurves are shown in Figures 1 and 2. The lightcurve amplitude from the January observations is around 0.045 mag. over the rotational phase observed. The derived mean apparent magnitude is $R = 16.551 \pm 0.008$. Using $G = -0.07 \pm 0.02$ and $V - R = 0.380 \pm 0.003$ from Pravec et al. (2000) we derive a mean absolute visual magnitude $H_V = 17.931 \pm 0.009$ using the $H, G$ system as defined by Bowell et al. (1989).

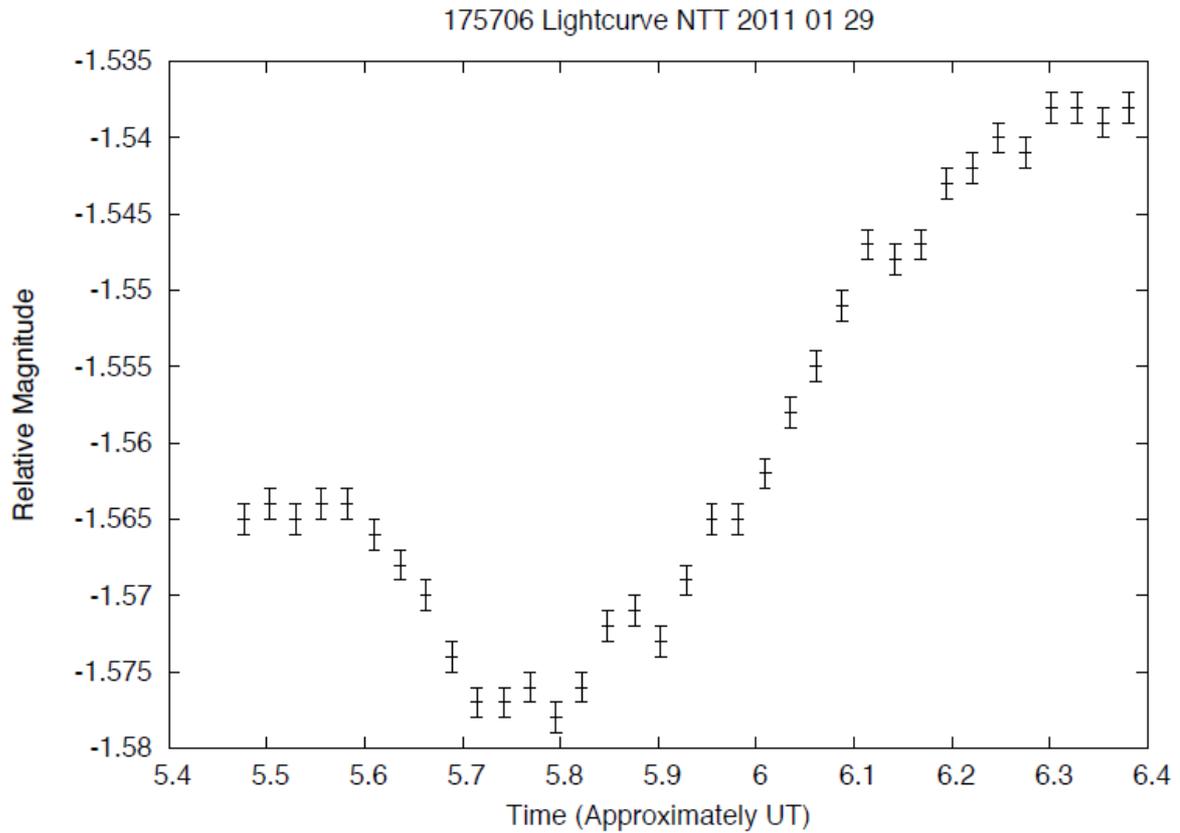

Fig. 1 Relative lightcurve observed on 29 January 2011.

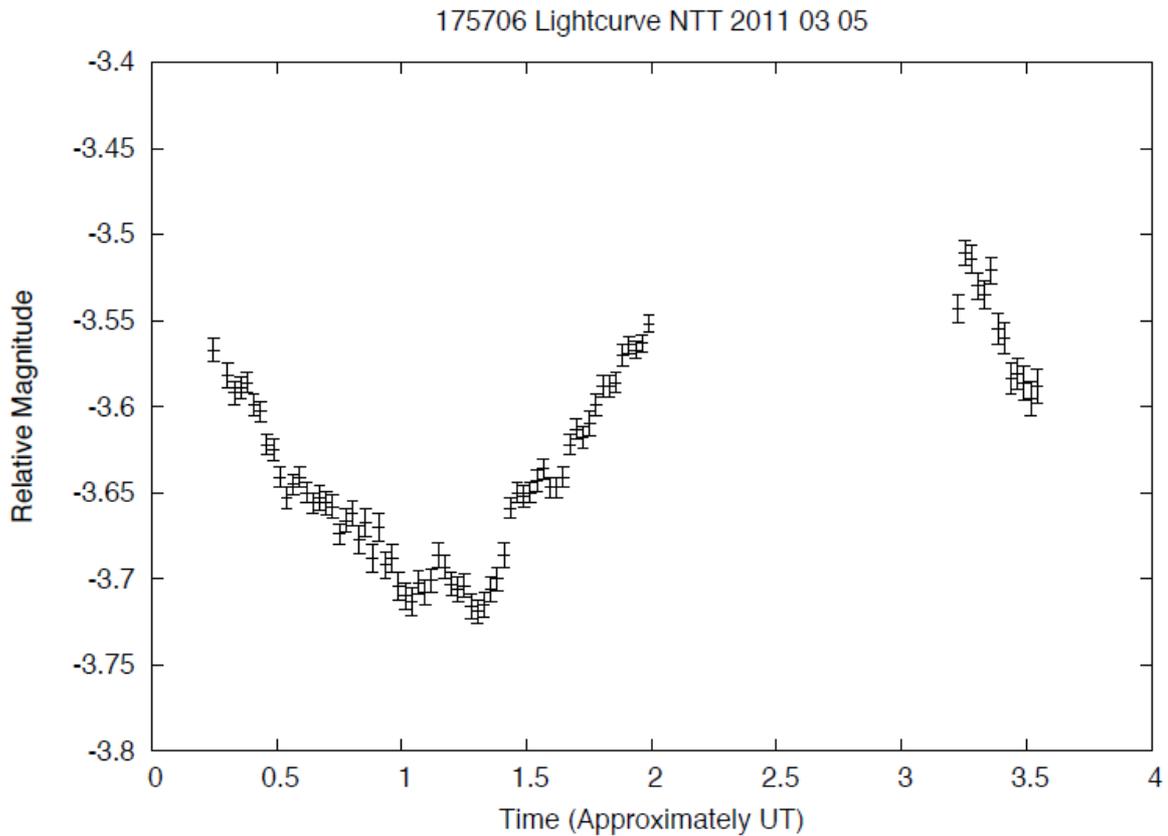

Fig 2. Relative lightcurve observed on 06 March 2011. The lightcurve amplitude appears much larger than the amplitude observed during January 2011.

The lightcurve observed in March has a significantly larger amplitude, >0.2 magnitudes. The derived mean apparent magnitude is $R$ = 18.942 ± 0.009 magnitudes. Using the same parameters as before we derive a mean absolute visual magnitude $H_V$ = 17.788 ± 0.009.

### 2.1.2 H, G analysis

All previous photometry on this asteroid was acquired at phase angles > 14°, with no sampling over a phase angle range including the value of the VLT observations (11.7°). Our optical observations at a phase angle of 1.4° are particularly important for characterizing the phase darkening behaviour over a much broader range of phase angles, including the important opposition surge region, and for a more robust measurement of the absolute magnitude of this object.

We combined our data with that in Pravec et al. (2000) and Mottola & Lahulla (2000), and conducted a new analysis of the phase curve over the range 1.4° - 47.5°. Pravec et al. cover a phase angle range of 13.9°- 32.1°, while Mottola & Lahulla cover 30.6° – 47.5°.

Reduced magnitudes from previous observations were extracted from Table 2 of Pravec et al. (2000) and from Figure 2 of Motolla & Lahulla (2000). For those reduced magnitudes extracted from Motolla & Lahulla, care was taken to avoid eclipse events and also to convert the reduced magnitudes to the phase angle at the time of the observations. This was accomplished by using the same phase angle coefficient, 0.044 mag./degree, used in that work.

Our March 2011 photometry at a phase angle of 34.3° is consistent with the earlier brightness measurements and this region of the phase curve (see Figure 3). However, as noted in Section 1.1, the observations were taken during an eclipse, and therefore we exclude this data point from our analysis. Our observations in January are reasonably consistent with the derived $G$ value of -0.07 from Pravec et al. We performed a chi-square minimization technique to find the new best fit values for $H$, $G$. Our new fitted results for the entire dataset yield the following values: $H_R$ = 17.453 ± 0.012 mag., $G$ = -0.041 ± 0.005.

Figure 4 shows the results of this analysis along with the formal 1-3 sigma uncertainty regions as defined in Press et al. (1992) and Wall & Jenkins (2003). The best fit $H$, $G$ curve is plotted in Figure 3, along with the 3-sigma uncertainty curves.
However, the H, G function that is used for representing the asteroid's phase relation is not a perfect model (Harris, 1991). A more complete model is generally needed to describe the phase relation accurately; such models have parameters that cannot be solved for with only one data point at low solar phases and therefore cannot be used in this case. The H, G function may have a systematic model uncertainty of a few 0.01 mag; we assign 0.02 mag. as the uncertainty contributed. This leads to $H_V$ = 17.833 ± 0.024 magnitudes, using $V - R$ = 0.380 ± 0.003 from Pravec et al. The new absolute magnitudes can then be used to derive albedo values through the application of Eq. 1.

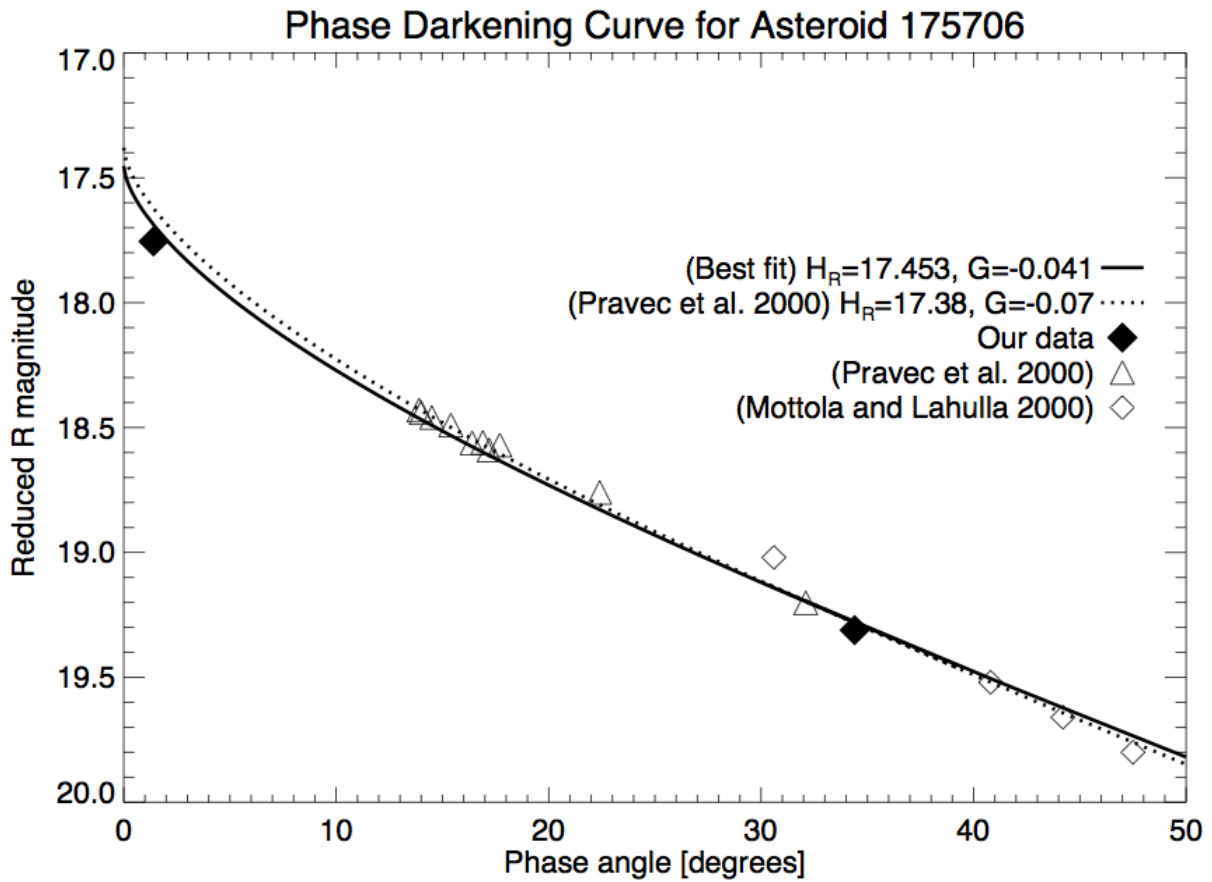

Fig 3. R-band phase darkening curve for 1996 FG3, combining our observations with those identified in the legend. Note that the 34.3° point is excluded from the fit.

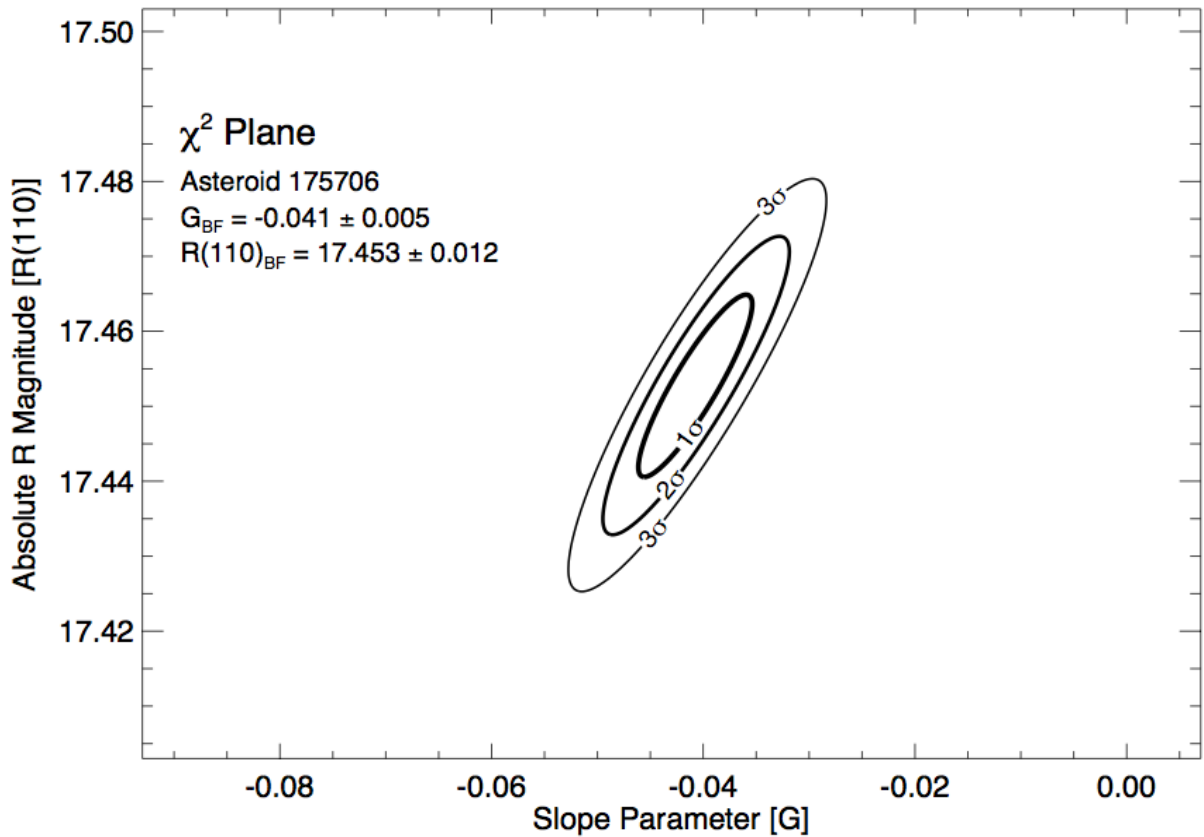

Fig. 4. Results of chi-squared minimisation technique for determining $H$, $G$.

## 2.2 Mid-IR Photometry and Simple Thermal Modelling

The VISIR intermediate field was used, which has a pixel scale of 0.127 arcsec per pixel and a total field-of-view of 32.5" × 32.5". The telescope was tracked at a rate of around -0.033 arcsec/s (dRA × $cos$ DEC)/dt and 0.0071 arcsec/s dDEC/dt on top of the sidereal rate (readjusted every hour). The observations were perpendicular chop-nodded with a throw of 8''. The filters used are shown in Table 2.

**Table 2: VISIR filters and exposure times used**

| Filter | Central Wavelength $\lambda_c$ (μm) | Detector Integration Time (DIT) (s) | Total exposure time of nod pair (s) | Estimated* / Advertised sensitivity (mJy/10σ/1h) |
|---|---|---|---|---|
| J8.9 | 8.70 | 0.0250 | 179.40 | 10.3 / 5 |
| B10.7 | 10.65 | 0.0100 | 181.24 | 12.4 / 8 |
| B11.7 | 11.52 | 0.0200 | 180.32 | 8.1 / 6 |
| B12.4 | 12.47 | 0.0125 | 180.55 | 16.2 / 12 |

* From standard observations

The standard stars HD26967, HD12524 and HD61935 were selected from Cohen et al. (1999). They were observed at three different airmasses covering a similar range to the asteroid (1.0-1.9) and within around 2hr RA. An atmospheric extinction correction (< 5%) and aperture correction (13-15%) was applied based on the calibration obtained. The typical FWHM and estimated uncertainties are given in Table 3.

**Table 3: Extinction and Aperture Correction Uncertainties**

| Filter | Extinction Correction Uncertainty (mag.) | Typical FWHM of standard PSF (pixels) | Aperture correction uncertainty (%) |
|---|---|---|---|
| J8.9 | 0.02 | 2.5 | 3 |
| B10.7 | 0.01 | 2.7 | 2 |
| B11.7 | 0.01 | 2.7 | 2 |
| B12.4 | 0.02 | 2.9 | 2 |

Every nod pair was combined onto a single frame using the VISIR pipeline, which removes stripes caused by detector instabilities (Pantin, Vanzi & Weilenman 2008). The observations were reduced with 3, 5, 7 and 10 pixel apertures. Comparison of the calibrated fluxes measured through each aperture radius showed they were consistent, although the 5 pixel aperture was selected as it gave the highest S/N after incorporating the uncertainty contributions from photon statistics, extinction correction and aperture correction combined in quadrature. The flux from all four beams in each combined image was summed (after multiplying the negative beams by -1), using an aperture radius of 5 pixels and a background annulus 15-20 pixels from the Point Spread Function (PSF) centre. The observational circumstances and reduced fluxes are given in Table 4. Figure 5 shows the rotational phase of the fluxes compared to the January optical observations. No discernable thermal lighcurve can be identified with variation significantly larger than the observational uncertainties.

**Table 4: Reduced Mid-IR VISIR fluxes**

| MJD* - 2455580 (days) | Average airmass | $\lambda_c$ (µm) | Flux (W m$^{-2}$ µm$^{-1}$) x 10$^{15}$ | Uncertainty (W m$^{-2}$ µm$^{-1}$) x 10$^{15}$ |
|---|---|---|---|---|
| 0.14598 | 1.693 | 11.52 | 4.91 | 0.17 |
| 0.14838 | 1.671 | 11.52 | 4.63 | 0.16 |
| 0.15077 | 1.650 | 11.52 | 4.76 | 0.16 |
| 0.15473 | 1.618 | 8.70 | 4.96 | 0.28 |
| 0.15725 | 1.599 | 8.70 | 4.78 | 0.26 |
| 0.15966 | 1.581 | 8.70 | 5.41 | 0.28 |
| 0.16884 | 1.520 | 11.52 | 5.07 | 0.19 |
| 0.17124 | 1.506 | 11.52 | 4.89 | 0.17 |
| 0.17852 | 1.468 | 10.65 | 5.01 | 0.25 |
| 0.18108 | 1.456 | 10.65 | 5.92 | 0.26 |
| 0.18346 | 1.445 | 10.65 | 5.34 | 0.25 |
| 0.18930 | 1.420 | 11.52 | 5.07 | 0.17 |
| 0.19175 | 1.411 | 11.52 | 5.14 | 0.17 |
| 0.19467 | 1.400 | 12.47 | 4.48 | 0.24 |
| 0.19711 | 1.392 | 12.47 | 4.26 | 0.23 |
| 0.19949 | 1.385 | 12.47 | 4.70 | 0.24 |
| 0.20183 | 1.377 | 12.47 | 4.48 | 0.24 |
| 0.20421 | 1.371 | 12.47 | 4.68 | 0.24 |
| 0.20668 | 1.364 | 12.47 | 4.15 | 0.23 |
| 0.20962 | 1.357 | 11.52 | 4.43 | 0.16 |
| 0.21205 | 1.352 | 11.52 | 5.02 | 0.17 |
| 0.21575 | 1.344 | 8.70 | 5.55 | 0.30 |
| 0.21844 | 1.339 | 8.70 | 5.52 | 0.28 |
| 0.22179 | 1.334 | 8.70 | 5.30 | 0.27 |
| 0.22856 | 1.325 | 11.52 | 5.13 | 0.17 |
| 0.23097 | 1.322 | 11.52 | 4.91 | 0.16 |
| 0.23438 | 1.319 | 10.65 | 5.46 | 0.26 |
| 0.23701 | 1.318 | 10.65 | 5.43 | 0.25 |
| 0.23937 | 1.317 | 10.65 | 5.63 | 0.25 |
| 0.24282 | 1.315 | 11.52 | 5.13 | 0.17 |
| 0.24525 | 1.315 | 11.52 | 5.31 | 0.17 |
| 0.25203 | 1.316 | 12.47 | 4.74 | 0.23 |
| 0.25445 | 1.317 | 12.47 | 4.53 | 0.23 |
| 0.25683 | 1.318 | 12.47 | 4.41 | 0.23 |

* Mean Julian Day mid-observation, light-time corrected.

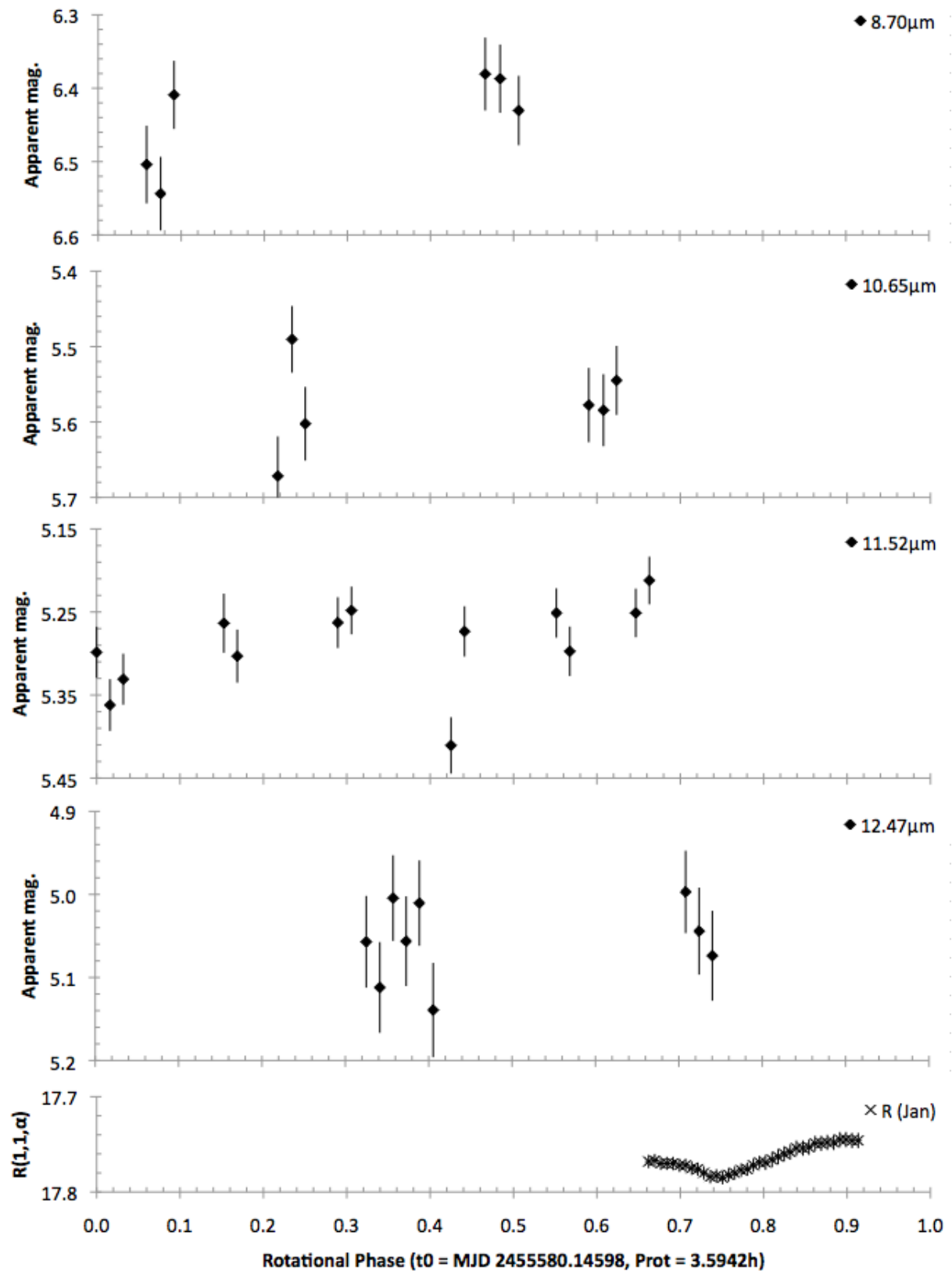

Fig. 5. Thermal-IR and R-band January observations of 1996 FG3, corrected to mid-observation time and light-time corrected, comparing the rotational phases. X-axis error bars on the R-band observations reflect the accumulated uncertainty in rotational phase since the time of thermal-IR observations, resulting from the 0.0002 h uncertainty in the rotational period. Zero mag fluxes for thermal IR magnitudes are derived from values given by Tokunaga (2000) and Beckwith et al. (1976). 8.70 μm, 10.65 μm and 12.47 μm observations were interspersed with 11.52 μm, which can be used to assess the rotational thermal IR lightcurve variability. NEATM fitting after lightcurve correcting with either a linear fit to the 11.52 micron

fluxes, or a linear interpolation between the fluxes, altered the derived albedo by <0.001, and so is neglected.

The derived simple model diameter/albedos are given in Table 5, with the fits shown in Figure 6, where it can be seen that the NEATM fits well to the observed fluxes, and the STM and FRM less so. The estimated uncertainty of the NEATM fit is ~15% in diameter and ~30% in albedo, dominated by the contribution of uncertainty from simple-model assumptions (see Discussion).

**Table 5: Simple thermal model derivations of diameter/albedo**

| Model | $p_v$ | $D_{eff}$ (km) | $\eta$ |
|---|---|---|---|
| STM | 0.059 | 1.42 | |
| FRM | 0.018 | 2.57 | |
| NEATM | 0.046 ± 0.014 | 1.68 ± 0.25 | 1.15 |

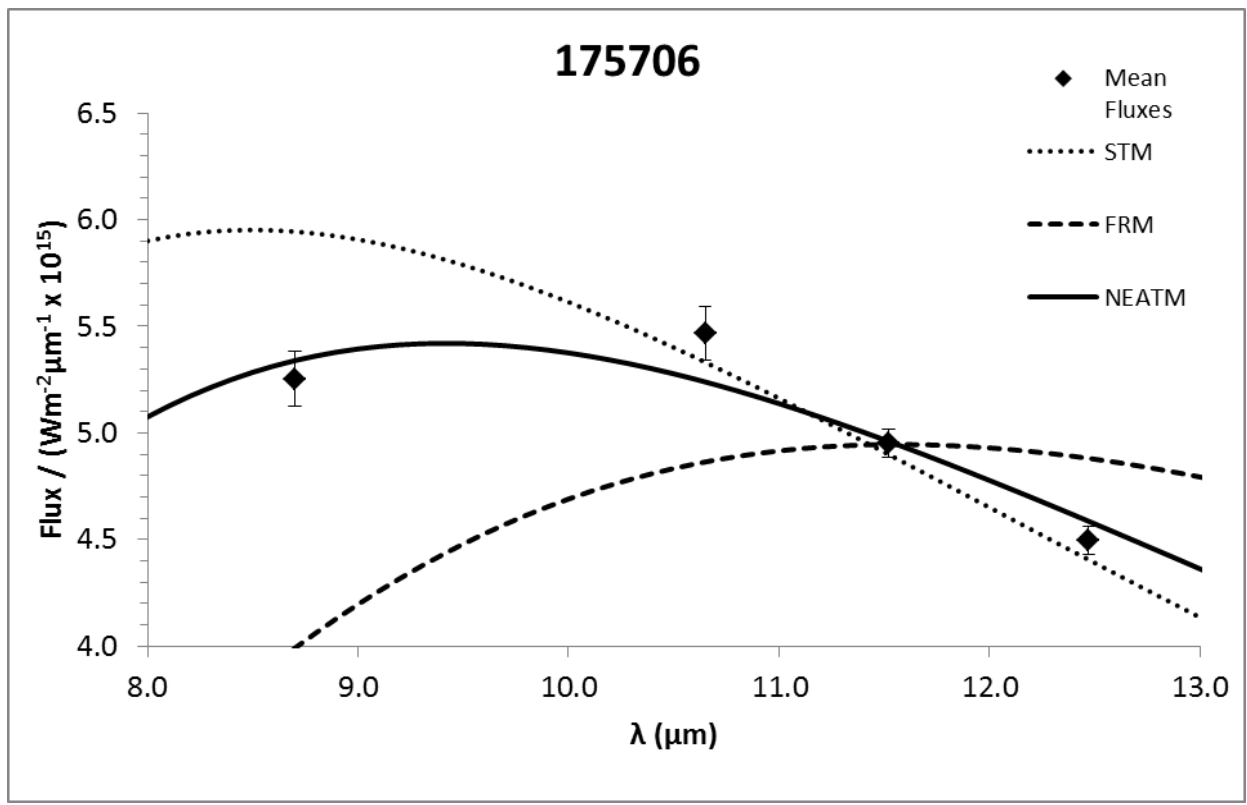

Fig. 6. STM, FRM, and NEATM fits to the thermal-IR observations of 1996 FG3. Although mean fluxes are plotted for clarity, fits were to all the fluxes (error-weighted) as given in Table 4. Error bars are the standard error of the observed fluxes at each wavelength.

### 2.3 Thermophysical Modelling

There is sufficient observational information about 1996 FG3, as outlined in Sections 1 and 2.1, to construct a thermophysical model for more detailed interpretations of the thermal IR observations. Here the ATPM, described in detail in Rozitis & Green (2011), is the thermophysical model used to fit the observations. The binary system is treated as two separate independent bodies to simplify the thermophysical modelling process of a two-period system. Fortunately, the VLT observations were taken outside of a mutual event, so the two components can be treated as independent objects in the modelling. This neglects mutual self-heating between the two components but it can be safely ignored since the maximum subtended sky fractions are ~0.005 for the secondary as observed by the primary at its sub-secondary point, and ~0.05 for the primary as observed by the secondary at its sub-primary point (i.e. the associated radiative heat transfer viewfactors are negligible). The shapes and rotation states used are those determined by Scheirich & Pravec (2009). In particular, the primary shape has a mean equatorial to polar radius ratio of 1.2, and the secondary shape is a prolate spheroid with a long to short axis ratio of 1.4. Unlike Scheirich & Pravec (2009), we adopt a triaxial ellipsoid shape rather than an oblate spheroid shape for the primary, to ensure that a rotational lightcurve can be produced for it. The primary shape is given a long to short equatorial axis ratio that is consistent with a maximum lightcurve amplitude of 0.08 magnitudes (Pravec et al. 2000) (hence would produce a lightcurve amplitude of 0.02 when in the geometry of the January observations). The adopted shape for the primary is consistent with that determined by Mottola & Lahulla (2000). We also assume the secondary to primary effective diameter ratio at the equatorial aspect to be 0.30 ± 0.01 as determined by the updated model produced by Pravec and collaborators (P. Pravec, pers. comm.).

Both component shape models are represented by triangular facet meshes consisting of 614 vertices and 1224 facets. Like Mottola & Lahulla (2000) and Scheirich & Pravec (2009), the rotation pole orientation of both components is assumed to be perpendicular to the binary system orbital plane and with the same sense of rotation. The secondary rotation is assumed to be tidally locked with its orbit about the primary and with its long axis aligned with the line drawn between the two centres of the bodies. Radar observations of binary NEA 1999 KW4 show that this is a likely configuration for binary asteroids (Ostro et al. 2006, Scheeres et al. 2006). Table 6 summarises the system properties adopted in the thermophysical modelling.

**Table 6: Assumed thermophysical modelling parameters**

| Property | Value |
|---|---|
| Primary shape ($A_p:B_p:C_p$) | 1.245:1.157:1.000 |
| Primary rotation period | 3.5942 hours |
| Secondary shape ($A_s:B_s:C_s$) | 1.4:1.0:1.0 |
| Secondary rotation period | 16.14 hours |
| Secondary to primary effective diameter ratio | 0.30 |
| Orbital plane and rotation pole orientation | $\beta = -84°, \lambda = 242°$ |
| Emissivity | 0.9 |
| Number of vertices | 614 (×2) |
| Number of facets | 1224 (×2) |

The free parameters to be constrained by fits to the thermal-IR observations are therefore the effective size at the observed aspect, albedo, thermal inertia, and surface roughness. The effective size and albedo are related by Equation 1 and can be considered a single free parameter. Surface roughness is represented by a fractional coverage, $f_R$, of hemispherical craters with the remaining fraction, 1 - $f_R$, representing a smooth flat surface. The hemispherical crater model used is the 132-facet type introduced in Rozitis & Green (2011), which was shown to accurately reproduce the lunar thermal-IR beaming effect. Surface roughness is also linked to the asteroid size and albedo, since it decreases the effective Bond albedo of a rough surface relative to a smooth flat one. For the hemispherical crater surface roughness representation, the effective Bond albedo, $A_{B\_EFF}$, is given by

$$A_{B\_EFF} = f_R \frac{A_B}{2-A_B} + (1 - f_R)A_B \qquad (3)$$

where $A_B$ is the Bond albedo of a smooth flat surface. Thus each size and roughness fraction combination leads to a unique Bond albedo value to be used in the ATPM. However, it is computationally expensive to run a separate thermophysical model for each Bond albedo value. Since the NEATM fit to the thermal-IR observations revealed a very low $p_v$ value, and hence very low Bond albedo, it is possible to run ATPM for one Bond albedo and perform a pseudo-correction to the predicted observed flux for a different Bond albedo. This flux correction factor, $FCF$, is given by

$$FCF = \frac{1-A_B}{1-A_{B\_MOD}} \qquad (4)$$

where $A_B$ is the smooth surface Bond albedo calculated by inversion of Equation 3, and $A_{B\_MOD}$ is the Bond albedo used in the ATPM. A model Bond albedo of 0.01 is assumed for both the primary and secondary, which implies that they also have the same $p_v$ values. Lightcurve fits to eclipses and occultations of binary asteroids by Pravec et al. (2006) show that the albedos of the two components in a binary system never differ by more than 20% making this a valid assumption. As the albedo is very low, most of the flux correction factors used in the thermophysical modelling process were very near unity.

A range of thermal inertias was chosen and a thermophysical model was run for each thermal inertia value for both the primary and secondary. To determine the thermal emission, the ATPM computes the surface temperature variation for each shape and roughness facet during a rotation by solving the 1D heat conduction equation with a surface boundary condition that includes direct and multiple scattered solar radiation, shadowing, and reabsorbed thermal radiation from interfacing facets. A Planck function is applied to the derived temperatures and summed across visible facets to give the emitted flux as a function of observation wavelength and roughness fraction. The respective fluxes from the primary and secondary were scaled according to their size ratio and then summed to give the overall flux. Since the rotational phase of the primary at the time of the observations was unknown, the primary flux was rotationally averaged. The overall model flux predictions, $F_{MOD}(\lambda_n, \Gamma, D_{eff}, f_R)$, were compared with the observations, $F_{OBS}(\lambda_n)$, and observational errors, $\sigma_{OBS}(\lambda_n)$, by varying the model combined effective diameter, $D_{eff}$, and roughness fraction, $f_R$, to give the error-weighted least squares fit

$$L^2 = \sum_{n=1}^{N} \left( \frac{FCF(D_{eff}, f_R)F_{MOD}(\lambda_n, \Gamma, D_{eff}, f_R) - F_{OBS}(\lambda_n)}{\sigma_{OBS}(\lambda_n)} \right)^2 \qquad (5)$$

for a set of *N* observations.

Table 7 summarises the model-fitting for the combined effective diameters that gave the minimum error-weighted least squares fit for some of the surface properties considered; Figure 7 displays an example model fit. Inspection of Table 7 indicates that the observations place a strong constraint on the thermal inertia but not on the surface roughness. The lack of constraint on the degree of surface roughness is expected because of the single phase angle at which the observations were conducted. As demonstrated in Rozitis & Green (2011), multiple observations at phase angles over a large range is required to constrain the degree of surface roughness.

As indicated by Table 7, a range of thermal inertias and roughness fractions give similar error-weighted least squares values. In order to determine the average values and the uncertainties of the ATPM fitted parameters, a fitting procedure was developed that found all the model variants that gave an error-weighted least squares fit that fell within 10% of the minimum value. The reported best-fit parameters are the average values from the accepted model variants and the uncertainties are the standard deviation. It also took into account the uncertainties in the derived *H* and *G* values by producing a randomly selected range of values, for use by the model variants, that had normal distributions with means and widths equal to the derived values and their uncertainties respectively. We note that the constant of 1329 km in Eq. 1 was derived using an estimate for the apparent visual magnitude of the Sun of $V_{SUN}$ = -26.762 ± 0.017 (Campins, Rieke & Lebofsky, 1985) which propogates into an uncertainty for the constant of ± 10 km, as demonstrated by Pravec and Harris (2007). This uncertainty is accounted for by the fitting procedure also. A similar Monte Carlo analysis was applied to the thermal-IR flux measurements and their uncertainties, with the fitting procedure applied to each randomly generated spectra. However, it was found that the average results of the fits to the randomly generated spectra did not differ from fitting to the nominal unaltered spectra. This presumably happens because each wavelength was observed multiple times, giving multiple possible points within the measurement uncertainty for each wavelength anyway. We therefore quote the fitting procedure results to just the nominal unaltered spectra and they are summarised in Table 8. The combined effective diameter is converted to mean diameters of the primary and secondary using the shape parameters determined by Scheirich and Pravec (2009). The uncertainties for these diameters are larger than that for the combined effective diameter because of the large uncertainties associated with the derived shape parameters.

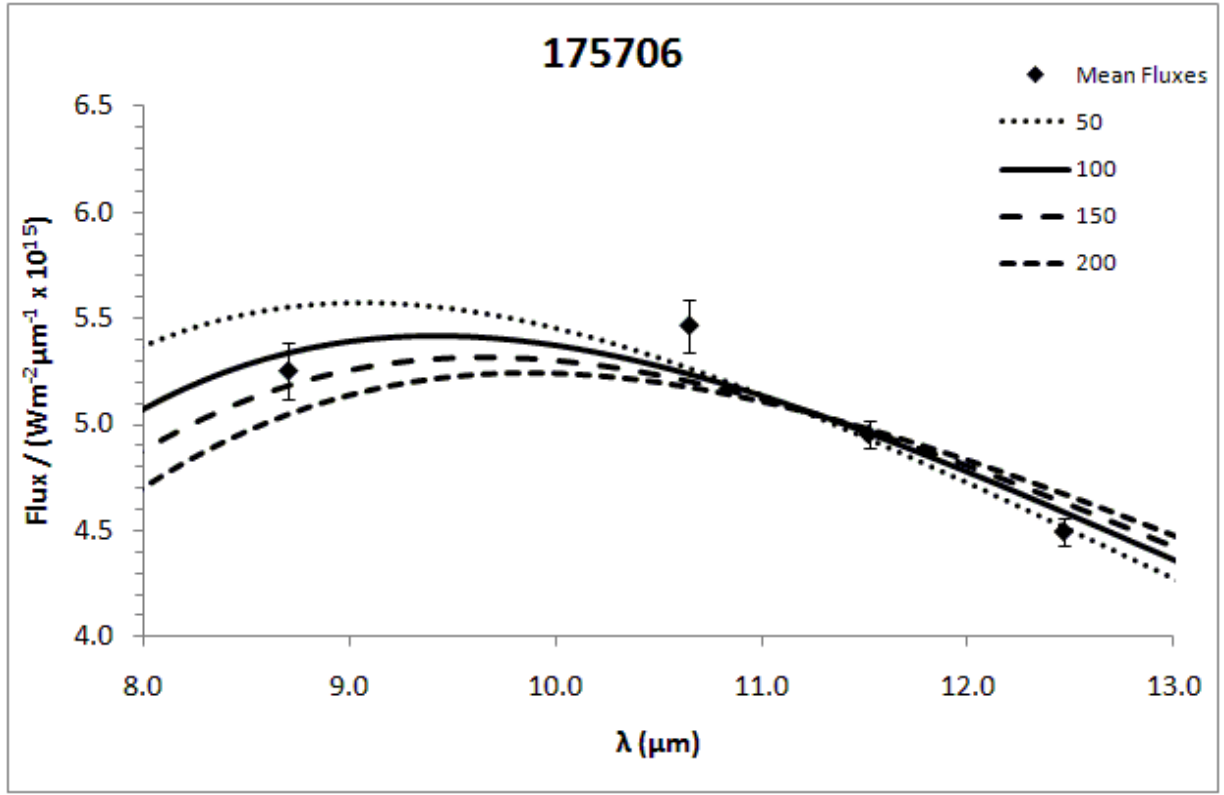

Fig. 7: ATPM fits to the thermal-IR observations of 1996 FG3, assuming $f_R = 0.4$, which corresponds to the level of surface roughness measured on the Moon (Rozitis & Green 2011). The different line styles correspond to different indicated levels of thermal inertia given in J m$^{-2}$ K$^{-1}$ s$^{-1/2}$. Although mean fluxes are plotted for clarity, fits were to all the fluxes (error-weighted) as given in Table 4.

**Table 7: ATPM fits to the thermal-IR observations of 1996 FG3**

| | | Thermal Inertia / J m$^{-2}$ K$^{-1}$ s$^{-1/2}$ | | | | | | | | | | | | | | |
|---|---|---|---|---|---|---|---|---|---|---|---|---|---|---|---|---|
| | | 0 | | 50 | | 100 | | 150 | | 200 | | 300 | | 400 | | 750 | |
| | | $L^2$ | Diameter | $L^2$ | Diameter | $L^2$ | Diameter | $L^2$ | Diameter | $L^2$ | Diameter | $L^2$ | Diameter | $L^2$ | Diameter | $L^2$ | Diameter |
| Roughness Fraction | 0 | 54.3 | 1587 | 52.4 | 1728 | 57.0 | 1843 | 65.3 | 1931 | 75.2 | 2010 | 93.3 | 2109 | 110.6 | 2192 | 151.5 | 2360 |
| | 0.05 | 56.2 | 1566 | 52.1 | 1712 | 55.7 | 1822 | 63.3 | 1916 | 72.9 | 1994 | 90.7 | 2093 | 107.8 | 2177 | 148.3 | 2344 |
| | 0.10 | 58.3 | 1545 | 52.1 | 1691 | 54.7 | 1806 | 61.7 | 1900 | 70.7 | 1973 | 88.2 | 2083 | 105.0 | 2166 | 145.3 | 2328 |
| | 0.15 | 60.6 | 1524 | 52.5 | 1676 | 53.9 | 1785 | 60.1 | 1879 | 68.6 | 1958 | 85.8 | 2067 | 102.4 | 2151 | 142.3 | 2318 |
| | 0.20 | 63.2 | 1503 | 52.9 | 1655 | 53.1 | 1770 | 58.7 | 1864 | 66.8 | 1942 | 83.6 | 2052 | 100.0 | 2135 | 139.5 | 2302 |
| | 0.25 | 66.2 | 1488 | 53.5 | 1639 | 52.6 | 1754 | 57.5 | 1848 | 65.2 | 1926 | 81.6 | 2036 | 97.7 | 2125 | 136.8 | 2292 |
| | 0.30 | 69.0 | 1467 | 54.5 | 1623 | 52.3 | 1738 | 56.5 | 1832 | 63.7 | 1911 | 79.7 | 2025 | 95.4 | 2109 | 134.1 | 2276 |
| | 0.35 | 72.0 | 1451 | 55.5 | 1608 | 52.1 | 1723 | 55.7 | 1817 | 62.4 | 1900 | 77.8 | 2010 | 93.4 | 2093 | 131.7 | 2266 |
| | 0.40 | 75.2 | 1436 | 56.5 | 1592 | 52.1 | 1707 | 54.9 | 1806 | 61.0 | 1884 | 76.2 | 1999 | 91.3 | 2083 | 129.2 | 2250 |
| | 0.45 | 78.4 | 1420 | 57.7 | 1576 | 52.3 | 1691 | 54.1 | 1791 | 59.9 | 1869 | 74.5 | 1984 | 89.5 | 2072 | 126.9 | 2239 |
| | 0.50 | 81.6 | 1404 | 58.9 | 1561 | 52.3 | 1681 | 53.7 | 1775 | 59.0 | 1858 | 73.0 | 1973 | 87.7 | 2057 | 124.7 | 2224 |
| | 0.55 | 84.8 | 1389 | 60.3 | 1545 | 52.5 | 1665 | 53.2 | 1764 | 57.9 | 1843 | 71.7 | 1958 | 85.9 | 2046 | 122.5 | 2213 |
| | 0.60 | 88.2 | 1373 | 61.8 | 1535 | 53.0 | 1655 | 52.8 | 1749 | 57.2 | 1832 | 70.2 | 1947 | 84.3 | 2036 | 120.4 | 2203 |
| | 0.65 | 91.5 | 1362 | 63.1 | 1519 | 53.3 | 1639 | 52.5 | 1738 | 56.3 | 1817 | 69.0 | 1937 | 82.8 | 2020 | 118.4 | 2192 |
| | 0.70 | 94.7 | 1347 | 64.8 | 1509 | 53.8 | 1629 | 52.4 | 1723 | 55.6 | 1806 | 67.9 | 1926 | 81.3 | 2010 | 116.5 | 2182 |
| | 0.75 | 98.1 | 1336 | 66.3 | 1493 | 54.4 | 1613 | 52.1 | 1712 | 55.1 | 1796 | 66.8 | 1916 | 79.8 | 1999 | 114.6 | 2166 |
| | 0.80 | 101.5 | 1321 | 67.8 | 1483 | 54.8 | 1603 | 52.0 | 1702 | 54.6 | 1780 | 65.7 | 1900 | 78.5 | 1989 | 112.8 | 2156 |
| | 0.85 | 104.6 | 1310 | 69.5 | 1472 | 55.4 | 1592 | 52.0 | 1691 | 54.0 | 1770 | 64.7 | 1890 | 77.2 | 1978 | 111.1 | 2145 |
| | 0.90 | 107.7 | 1300 | 71.2 | 1462 | 56.1 | 1582 | 52.1 | 1681 | 53.6 | 1759 | 63.8 | 1879 | 75.9 | 1968 | 109.4 | 2135 |
| | 0.95 | 110.9 | 1289 | 72.9 | 1451 | 56.8 | 1571 | 52.2 | 1670 | 53.2 | 1749 | 62.9 | 1869 | 74.8 | 1958 | 107.8 | 2125 |
| | 1 | 114.1 | 1279 | 74.5 | 1441 | 57.6 | 1561 | 52.3 | 1660 | 52.9 | 1738 | 62.1 | 1864 | 73.7 | 1947 | 106.2 | 2119 |

Note: The dark, medium, and light gray cells indicate fits where the $L^2$ values are within 10%, 20%, and 30% of the minimum $L^2$ value respectively, and the combined effective diameters are given in metres.

**Table 8: 1996 FG3 properties derived by ATPM**

| Property | Value |
|---|---|
| Combined Effective Diameter, $D_{eff}$ (km) | 1.71 ± 0.07 |
| Primary Mean Diameter, $D_p$ (km) | 1.69 $^{+0.18}/_{-0.12}$ |
| Secondary Mean Diameter, $D_s$ (km) | 0.51 ± 0.03 |
| Geometric Visual Albedo, $p_v$ | 0.044 ± 0.004 |
| Thermal Inertia $\Gamma$ (J m$^{-2}$ K$^{-1}$ s$^{-1/2}$) | 120 ± 50 |

The ATPM fit assumed that both the primary and secondary have the same level of thermal inertia. However, they could potentially have different levels. To assess whether different levels of thermal inertia could be detected, the ATPM fitting was tested with the primary having low thermal inertia and the secondary having high thermal inertia, and vice versa. It was found that different primary and secondary thermal inertias did not significantly affect the fits, making it impossible to determine whether there is any thermal inertia difference. This indicates that the thermal-IR observations are only sensitive to the thermal inertia of the primary at the accuracy they were acquired. However, differing thermal inertias of the two bodies did give slightly different derived sizes. For the case of low thermal inertia for the primary and high thermal inertia for the secondary, the derived sizes were ~3% larger than the case of identical thermal inertia. The opposite case gave derived sizes that were ~4% smaller.

An assessment of the effect of hypothetical eclipses and occultations caused by the two binary components on the ATPM derived results was conducted, in order to examine the model's sensitivity to such occurrences. Two additional system configurations were considered: (1) secondary occulted (and in this specific geometry, eclipsed also) by the primary, and (2) secondary eclipsing and occulting the primary. The ATPM modelling in these two different configurations accounted for eclipse shadows where appropriate. In configuration (1), ATPM fitting derives a combined effective diameter $D_{eff}$ = 1.80 ± 0.07 km, an albedo $p_v$ = 0.040 ± 0.003, and thermal inertia $\Gamma$ = 110 ± 50 J m$^{-2}$ s$^{-1/2}$ K$^{-1}$. This configuration gives a combined effective diameter ~5% larger and albedo ~9% smaller than the separated side-by-side configuration because of the lack of extra cross-sectional area projected towards the observer provided by the secondary. In configuration (2), ATPM fitting derives a combined effective diameter $D_{eff}$ = 1.83 ± 0.07 km, an albedo $p_v$ = 0.039 ± 0.003, and thermal inertia $\Gamma$ = 90 ± 50 J m$^{-2}$ s$^{-1/2}$ K$^{-1}$. This configuration gives a combined effective diameter ~7% larger and albedo ~11% smaller because of the presence of both an eclipse shadow and the occultation of the primary by the secondary. Although not knowing the secondary orbital position about the primary increases the uncertainty of the derived diameters and albedos, it is interesting to note that measuring the best-fit surface thermal inertia is not so sensitive to this knowledge.

## 3. Discussion

The derived low albedo $p_v$ = 0.044 ± 0.004 is consistent with the asteroid's primitive C-type spectroscopic classification. As Scheirich and Pravec (2009) found that the satellite has an orbital semi-major axis $a$ = (3.1 $^{+0.9}/_{-0.5}$) × $A_p$, we obtain $a$ = (2.8 $^{+1.7}/_{-0.7}$) km. Adopting the derived bulk density from Scheirich & Pravec of (1.4 $^{+1.5}/_{-0.6}$) g cm$^{-3}$, we determine a mass of the primary $M_p$ = (3.5 $^{+6.4}/_{-1.9}$) × 10$^{12}$ kg and secondary $M_s$ = (1.0 $^{+1.4}/_{-0.5}$) × 10$^{11}$ kg. (

Mueller et al. (2011), as part of the ExploreNEOs Warm-Spitzer Survey, observed 1996 FG3 on 2 May 2010. They measured $D_{eff}$ = 1.84 ($^{+0.56}/_{-0.47}$) km and $p_v$ = 0.042 $^{+0.035}/_{-0.017}$ which is in excellent agreement with our results. Although Warm-Spitzer observes in two channels with central wavelengths of 3.6 and 4.5 μm, the reliability of the 3.6 μm channel is questionable as the reflected flux has to be removed. The assumption is made that the spectral reflectivity is 1.4 times that in the V-band, and in rare cases this can lead to negative calculated 3.6 μm fluxes. Hence, a beaming parameter cannot be reliably derived by fitting to the two fluxes. Mueller et al. applied a fixed "beaming parameter" $\eta$ using an assumed linear empirical relationship between $\eta$ and the phase angle $\alpha$ = 51.02° (Delbo' et al. 2003; Wolters et al. 2008):

$$\eta = (0.91 \pm 0.17) + (0.013 \pm 0.004) \propto °  \qquad (6)$$

from which we infer that $\eta$ = 1.57 was applied. Mueller et al. also assumed $H_V$ = 17.76 ± 0.03 and $G$ = -0.07 from Pravec et al. (2000). The uncertainty contribution from assuming a fixed $\eta$ is robustly estimated, along with the measured flux, calibration and $H_V$ uncertainties, with a Monte-Carlo approach, in which 1000 sets of random synthetic fluxes are normally distributed about the measured value with a standard deviation equal to the root-sum-square of these uncertainties. It was assumed that the true beaming parameter $\eta$ was within ±0.3 of the true value. This uncertainty estimate is corroborated by Ryan & Woodward (2010) who found a typical $\eta$ value of 1.07 ± 0.27. However, Mueller et al. caution that this sample is dominated by large main-belt asteroids. More work needs to be done to ascertain the reliability of this assumption for small NEAs.

Best-fitting the beaming parameter attempts to compensate for an altered temperature distribution due to beaming. It can also compensate for, to some extent, non-zero thermal inertia and not including thermal emission from the night side. A study that compares NEATM fits with a thermophysical model and radar-derived diameters (Wolters & Green 2009) shows that even when able to fit $\eta$, NEATM systematically underestimates diameter and overestimates albedo. At phase angles $\alpha$ > 45°, NEATM can become significantly inaccurate, and will overestimate the diameter by ~10%-40% and underestimate the albedo. The inaccuracy increases with $\alpha$ and depends on whether the morning or evening hemisphere is observed, i.e. it is dependent on whether the asteroid rotates in a prograde or retrograde sense. The assessment of uncertainty in the ExploreNEOs survey does not currently account for these effects.

A prediction of the Yarkovsky effect acting on 1996 FG3 can be made using the derived shape and range of likely surface thermal properties. Unfortunately it is not possible to make a YORP effect prediction since the shape model used is a triaxial ellipsoid, which is not susceptible to YORP torques. The Yarkovsky effect can be determined by computing the total recoil force from photons thermally emitted from the shape and roughness facets (Rozitis & Green 2010, submitted to MNRAS). Since 1996 FG3 is on a highly elliptical heliocentric orbit and has a pole orientation not perpendicular to its orbital plane, an accurate prediction is made by averaging the Yarkovsky forces both over the asteroid rotation and its orbit. We predict the value of this effect acting on both the primary and secondary by averaging over one rotation at 16 orbital points equally spaced by 22.5° in true anomaly. We find that for a bulk density of (1.4 $^{+1.5}/_{-0.6}$) g cm$^{-3}$, a thermal inertia of 120 J m$^{-2}$ K$^{-1}$ s$^{-1/2}$, and zero surface roughness (which implies a combined effective diameter of 1.87 km) gives a Yarkovsky-induced drift in its semi-major axis of (-50 $^{+26}/_{-37}$) m yr$^{-1}$ due to its retrograde rotation. For default $f_R$ = 0.4, as measured for the Moon (Rozitis and Green, 2011) (1.74 km size) the Yarkovsky drift is ($-60\ ^{+31}/_{-45}$) m yr$^{-1}$. For full surface roughness (1.60 km size), the Yarkovsky drift is ($-76\ ^{+39}/_{-57}$) m yr$^{-1}$. In these predictions, the secondary contributed ~6% of the total Yarkovsky force acting on the binary system. This predicted rate of Yarkovsky orbital drift is somewhat lower than that observed for (6489) Golevka ($-95.6 \pm 6.6$ m yr$^{-1}$, Chesley et al. 2003) and (152563) 1992 BF ($-160 \pm 10$ m yr$^{-1}$, Vokrouhlicky, Chesley & Matson 2008), but higher than that observed for (1862) Apollo ($-36.5 \pm 3.9$ m yr$^{-1}$, Chesley et al. 2008), making it a viable candidate for a future Yarkovsky detection.

This study provides the first measurement of thermal inertia for a binary NEA. 1996 FG3 has a rapidly rotating primary and close secondary, sharing these characteristics with ~32 out of the known 36 binary NEA systems (Delbo' et al, 2011). Delbo' et al. found that NEA binaries had an average high thermal inertia of $\Gamma = 480 \pm 70$ J m$^{-2}$ s$^{-1/2}$ K$^{-1}$, based on a statistical inversion method that fits to a distribution of $\eta$-$\alpha$ for a sample of synthetic objects with varying rotation period, subsolar latitude, sub-Earth latitude and longitude (Delbo' et al., 2007). This result is more than twice the mean thermal inertia found for NEAs ($\Gamma = 200 \pm 40$ J m$^{-2}$ s$^{-1/2}$ K$^{-1}$) found by Delbo' et al. (2007). Based on this result, Delbo' et al. (2011) concluded that the high thermal inertia provided support for a model of binary formation through YORP-induced rotational fission similar to that modelled by Walsh, Richardson & Michel (2008), as high thermal inertia indicates a bare rock surface at least partly stripped of regolith. Our derived thermal inertia $\Gamma = 120 \pm 50$ J m$^{-2}$ s$^{-1/2}$ K$^{-1}$ is inconsistent with this result, and is more consistent with a surface particle size distribution somewhat coarser, like that found on the Moon. This low value of thermal inertia suggests that the proposed regolith-sampling mechanism for MarcoPolo-R is appropriate for this target.

While it is impossible to draw conclusions about the typical surface properties of binary asteroids from one result, if the scenario described by Delbo' et al. is correct, it raises questions about the binary formation mechanism of 1996 FG3. However, even if 1996 FG3 was formed by YORP-induced rotational fission, there is no reason to assume that its residual surface would be solid rock. If the pre-binary object were a rubble pile then the concept of a regolith on a solid rock could be inappropriate. Even after regolith migration and equatorial loss, the surface could remain rough and retain a low thermal inertia.

## Acknowledgements

The work of S. D. Wolters and B. Rozitis was funded by the UK Science and Technology Facilities Council. The research leading to these results has received funding from the European Union Seventh Framework Programme (FP7/2007-2013) under grant agreement no. 268421. We thank P. Pravec for his thorough and insightful review and both Pravec and his collaborators for making it possible for us to refine our estimates through the applicaton of their preliminary updated orbital model.